\documentclass[aps,prl,twocolumn,superscriptaddress]{revtex4-2}
\usepackage[colorlinks,linkcolor=blue,anchorcolor=blue,citecolor=blue]{hyperref}
\usepackage{graphicx}
\usepackage{amsmath}
\usepackage{multirow}
\usepackage{float}

\begin{document}


\title{\textit{Ab Initio} Emergence and Collapse of Nuclear Collectivity near \textit{N}=\textit{Z}=40}


\author{X. C. Cao}
\affiliation{School of Physics and Astronomy, Sun Yat-sen University, Zhuhai, 519082, Guangdong, China}
\affiliation{Guangdong Provincial Key Laboratory of Quantum Metrology and Sensing, Sun Yat-sen University, Zhuhai, 519082, Guangdong, China}
\author{C. F. Jiao}
\email[]{jiaochf@mail.sysu.edu.cn}
\affiliation{School of Physics and Astronomy, Sun Yat-sen University, Zhuhai, 519082, Guangdong, China}
\affiliation{Guangdong Provincial Key Laboratory of Quantum Metrology and Sensing, Sun Yat-sen University, Zhuhai, 519082, Guangdong, China}
\author{R. Z. Hu}
\affiliation{School of Physics and State Key Laboratory of Nuclear Physics and Technology, Peking University, Beijing, 100871, China}


\date{\today}

\begin{abstract}
We present an \textit{ab initio} description of the emergence and collapse of enhanced quadrupole collectivity in nuclei near
$N=Z=40$ using multi-shell valence-space Hamiltonians derived from
chiral two- and three-nucleon forces with the in-medium similarity
renormalization group. Projected
generator-coordinate calculations with consistently evolved $E2$
operators capture the strong collectivity in $^{80}$Zr and its
decrease toward $^{86}$Mo and $^{88}$Ru without empirical effective
charges. While the $N,Z=40$ effective single-particle energy gaps remain open, the
$1g_{9/2}$--$2d_{5/2}$ spacing is smallest near $^{76}$Sr and
$^{80}$Zr, with its variation governed mainly by the proton-neutron
monopole contribution. The associated occupancies are
consistent with quadrupole correlations involving pseudo-SU(3)
$pf$ holes and quasi-SU(3) particles. Excluding the occupation of the proton and neutron $2d_{5/2}$ from the variational space strongly suppresses the deformation and $B(E2;2_1^+\rightarrow0_1^+)$ in $^{80}$Zr. These results mark a steady step toward \textit{ab initio} computations of heavy open-shell nuclei. 
\end{abstract}


\maketitle


{\it Introduction.}-- 
Explaining how nuclear collectivity emerges from underlying nuclear forces remains a central challenge in nuclear structure theory. This challenge is particularly acute in self-conjugate ($N=Z$) nuclei near $A=80$. The large $B(E2;2_1^+\rightarrow0_1^+)$ strengths observed in $^{76}$Sr and
$^{80}$Zr demonstrate the maximum of quadrupole collectivity along the $N=Z$ line~\cite{PhysRevLett.124.152501,PhysRevLett.49.308,PhysRevC.85.041303}. Recent lifetime measurements show
that $^{84}$Mo remains collective, whereas the $E2$ strengths are
substantially smaller in $^{86}$Mo and $^{88}$Ru~\cite{HaNC2025,r6ns-ypw8}. Precision mass measurements have also revealed enhanced binding in $^{80}$Zr, interpreted as evidence for a deformed double-shell closure at $N=Z=40$~\cite{HamakerNP2021}. These findings call for a fully microscopic analysis of how strong quadrupole correlations develop and weaken over a narrow range of proton and neutron numbers.

Energy density functional calculations describe the broad
evolution of collectivity and predict competing shapes in this region ~\cite{PhysRevLett.124.152501,PhysRevC.81.014303,RODRIGUEZ2011255,refId0}. Large-scale shell-model (LSSM) and discrete nonorthogonal shell-model
(DNO-SM) studies associate the enhanced collectivity with
quadrupole correlations involving pseudo-SU(3) $pf$ holes and
quasi-SU(3) particles~\cite{KANEKO2021136286,HaNC2025,r6ns-ypw8,10.1098/rspa.1958.0072,10.1098/rspa.1958.0101,PhysRevC.52.R1741,PhysRevC.92.024320}, and its reduction to an increasing $1g_{9/2}$--$2d_{5/2}$ spacings~\cite{HaNC2025,r6ns-ypw8}. All such studies, however, have been based on phenomenological or quasi-microscopic nuclear forces and density functionals; shell-model calculations also require empirical effective charges. Angular-momentum-projected coupled-cluster calculations~\cite{PhysRevC.105.064311,PhysRevX.15.011028,PhysRevC.111.044304} marked the first \textit{ab initio} computation to advance to this region~\cite{PhysRevC.110.L011302}, but a simultaneous
description of the rotational spectra and transition
strengths remains challenging~\cite{PhysRevC.110.L011302}. It therefore remains an open question whether the observed enhancement and subsequent weakening of collectivity in this region can be reproduced within a unified framework based on chiral nuclear forces.

Here we tackle this problem by combining multi-shell valence-space in-medium similarity renormalization group (VS-IMSRG) Hamiltonians~\cite{PhysRevLett.113.142501,HERGERT2016165,annurev:/content/journals/10.1146/annurev-nucl-101917-021120,PhysRevC.102.034320} with the projected generator-coordinate method (PGCM)~\cite{PhysRevC.96.054310,PhysRevC.98.064324,PhysRevC.110.054326,CAO2025140034}. We refer to this framework as VS-IM-GCM. Using a common chiral nucleon-nucleon (NN) and three-nucleon (3N) interaction and consistently evolved $E2$ operators, we study the even-even $N=Z$ nuclei from
$^{72}$Kr to $^{88}$Ru and the neighboring nucleus $^{86}$Mo.
We examine the underlying connections among collectivity, effective single-particle energy (ESPE) spacings, and cross-shell occupations, and test the role of the $2d_{5/2}$ orbit in $^{80}$Zr
by constraining its proton and neutron occupations to zero
in the intrinsic basis. No \textit{ad hoc} adjustments to the
interaction in this mass region or empirical effective charges are exploited.

\begin{figure*}[t]
  \includegraphics[width=\textwidth]{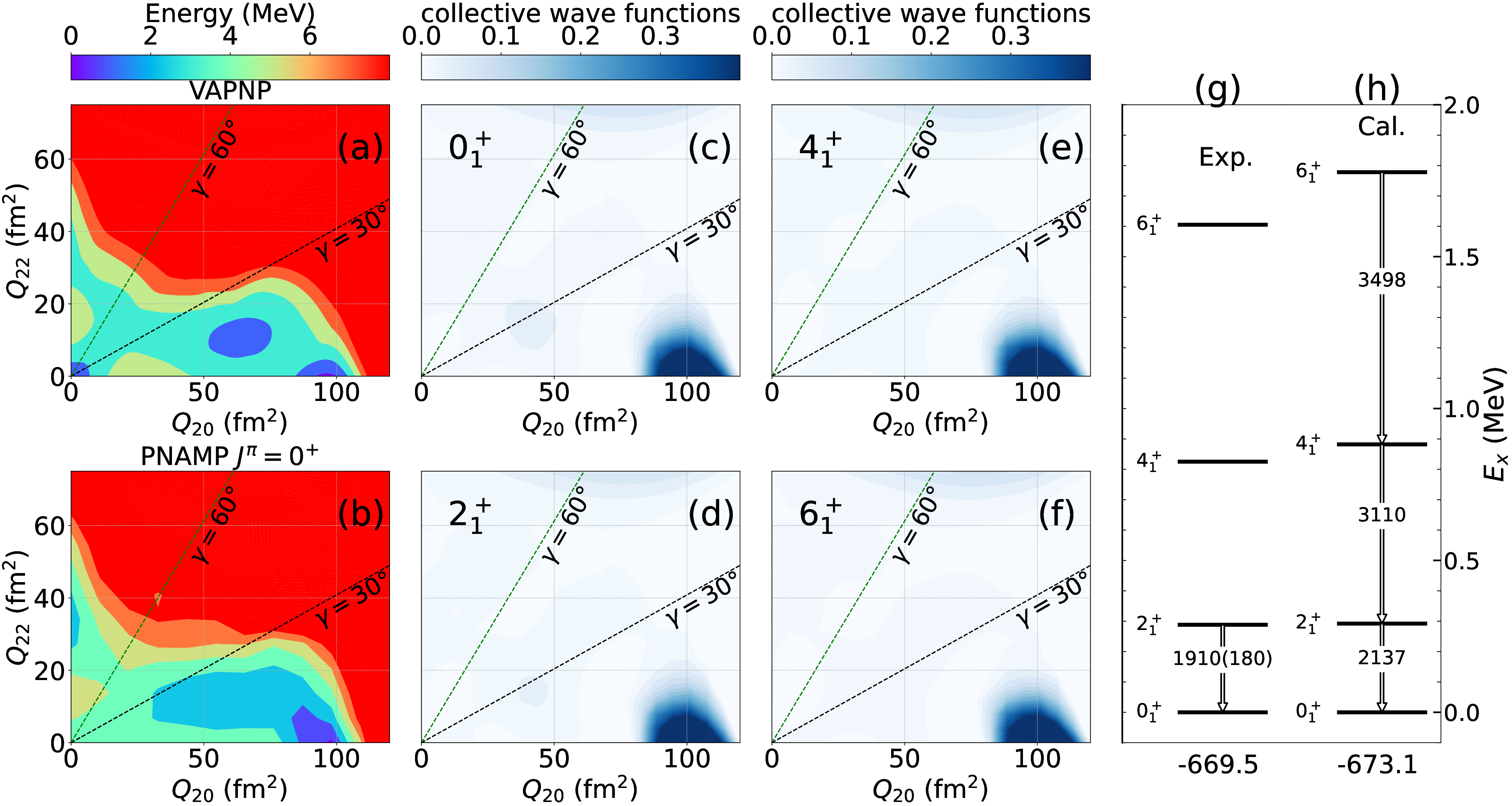}
  \caption{Calculated total energy surfaces, distributions of PGCM collective wave functions, experimental and calculated ground-state energies and low-lying spectra for $^{80}$Zr. The energy surfaces are (a) VAPNP and (b) projected to $N_{\text{valence}} = Z_{\text{valence}} = 12$ and $J^{\pi} = 0^+$. Each surface is measured from its own minimum. Panels (c)--(f) show the distributions of collective wave functions for $0_1^+$, $2_1^+$, $4_1^+$, and $6_1^+$ states, respectively. Ground-state energies are shown at the bottom of the level spectra (g) and (h). $B(E2\downarrow)$ values are given in $e^2\text{fm}^4$ and are indicated by arrows. Experimental data are taken from Refs.~\cite{HamakerNP2021,PhysRevLett.124.152501,ENSDF}.\label{Fig1}}
\end{figure*}

{\it Methods.}-- We derive the nucleus-dependent multi-shell effective Hamiltonian using the VS-IMSRG. The intrinsic Hamiltonian is built from the 1.8/2.0(EM) chiral interaction~\cite{PhysRevC.83.031301,PhysRevC.93.011302,PhysRevC.96.014303}, which consists of a free-space SRG-evolved NN interaction and a bare 3N force. This chiral interaction accurately reproduces the ground-state energies of nuclei up to $A\approx 100$~\cite{annurev:/content/journals/10.1146/annurev-nucl-101917-021120,PhysRevLett.120.152503,PhysRevLett.126.022501,PhysRevC.105.014302}. The full Hamiltonian is normal ordered with respect to an ensemble reference state determined for each nucleus~\cite{PhysRevLett.118.032502}. We then perform the VS-IMSRG unitary transformation on the Hamiltonian via the Magnus formulation~\cite{PhysRevC.92.034331}, using the arctangent type anti-Hermitian generator adopted in Ref.~\cite{PhysRevC.102.034320} with the energy denominator shift $\Delta=10$ MeV. It decouples the chosen core and active valence space from excitations outside the valence space in a nonperturbative way~\cite{HERGERT2016165,PhysRevC.102.034320}. We retain up to two-body operators in the continuous series of unitary transformations, known as the IMSRG(2) truncation~\cite{HERGERT2016165,PhysRevC.85.061304,PhysRevLett.106.222502,PhysRevLett.110.242501,PhysRevLett.120.152503}, to make the evolution computationally tractable.

We take the valence space composed of the $2p_{3/2}$, $1f_{5/2}$, $2p_{1/2}$, $1g_{9/2}$, $2d_{5/2}$, $1g_{7/2}$, $3s_{1/2}$, $2d_{3/2}$, and $1h_{11/2}$ proton and neutron orbits on top of a $^{56}$Ni core. It consists of all valence orbits between the 28 and 82 closed shells. No truncation based on an assumed SU(3) structure is imposed.
The VS-IMSRG decoupling is performed in a harmonic-oscillator basis spanning 13 major shells [$e_{\mathrm{max}} = \mathrm{max}(2n+\ell)=12$] with frequency $\hbar\omega=16$ MeV, and three-body matrix elements are restricted to $E_{3\mathrm{max}} =24$. To remove spurious center-of-mass (c.m.) excitations, we follow the prescription proposed in Refs.~\cite{PhysRevC.82.034330,PhysRevC.102.034320} by adding a Gl\"ockner-Lawson term $\beta H_{\text{c.m.}}$~\cite{GLOECKNER1974313} to the initial Hamiltonian at the beginning of the VS-IMSRG evolution. The oscillator frequency entering $H_{\mathrm{c.m.}}$ is set equal to the basis frequency. The dependence of the calculated low-lying spectra and $E2$ strengths on the $e_{\mathrm{max}}$, $E_{3\mathrm{max}}$ truncations and the scaling factor $\beta$ is examined in the Supplemental Material. These numerical checks do not include uncertainties from the input chiral interaction or omitted many-body correlations.

The electric quadrupole transition operator is evolved with the same
Magnus transformation as the Hamiltonian~\cite{PhysRevC.92.034331,PhysRevC.96.034324}. We retain its
one- and two-body terms at the IMSRG(2) level, thereby
including the corresponding renormalization from
configurations outside the valence space. No empirical effective charges are introduced.

Exact diagonalization in this space is impractical; we therefore solve the valence-space Hamiltonian with the PGCM as a variational approximation~\cite{PhysRevC.96.054310,PhysRevC.98.064324,PhysRevC.98.054311,PhysRevC.100.044308,PhysRevC.100.031303,PhysRevC.103.064302,PhysRevC.110.054326}. We generate intrinsic Bogoliubov vacua $|\Phi(q)\rangle$ by variation after particle-number projection (VAPNP)~\cite{ANGUIANO200262,PhysRevC.76.014308,PhysRevC.72.064303} constrained to quadrupole coordinates $q=(Q_{20},Q_{22})$. The intrinsic states preserve parity,
while the Bogoliubov transformation allows neutron-proton mixing. The neutron-proton correlations, e,g., the isoscalar pairing, can be developed variationally, although they are not treated as an additional generator coordinate due to the computational cost. We then construct many-body wave functions $|\Psi_{\sigma}^{JNZ}\rangle=\sum_{q,K} f_{q,\sigma}^{JK}|JMK;NZ;q\rangle$, where$\quad |JMK;NZ;q\rangle\equiv \hat {\mathcal{P}}_{MK}^J \hat{\mathcal{P}}^N \hat {\mathcal{P}}^Z |\Phi(q)\rangle$ and $\hat{\mathcal{P}}$'s denote the symmetry projection operators that project the intrinsic states onto states with good angular momentum ($J$) and the neutron ($N$) and proton ($Z$) numbers, respectively. We determine the weights $f_{q,\sigma}^{JK}$ by solving the Hill-Wheeler-Griffin variational equation~\cite{Ring1980}. 

The projected configurations form a nonorthogonal basis that may contain near-linear dependencies, resulting in numerical instabilities. We therefore follow an iterative prescription proposed in Refs.~\cite{Caurier1975,PhysRevC.105.054314}, which selects projected configurations iteratively according to their variational energy gains. The convergence with the retained projected basis is examined in the Supplemental Material.

{\it Results and discussion}.--
We first examine $^{80}\mathrm{Zr}$, where the large measured $E2$ strength and anomalous binding motivate a closer examination of its collective structure~\cite{PhysRevLett.124.152501,HamakerNP2021}. With the convention 
$\gamma=\arctan(\sqrt{2}Q_{22}/Q_{20})$ used here, the VAPNP energy surface in Fig.~\ref{Fig1}(a) exhibits three nearly degenerate coexisting minima corresponding to spherical, triaxial, and prolate shapes in the $(Q_{20},Q_{22})$ plane. This is consistent with the previous energy density functional study with the Gogny D1S interaction~\cite{RODRIGUEZ2011255}. Nevertheless, after particle-number and angular-momentum projection,  Fig.~\ref{Fig1}(b) reveals that the prolate minimum gains the largest symmetry-restoration energy and becomes energetically favored.

\begin{figure}[t]
  \includegraphics[width=\columnwidth]{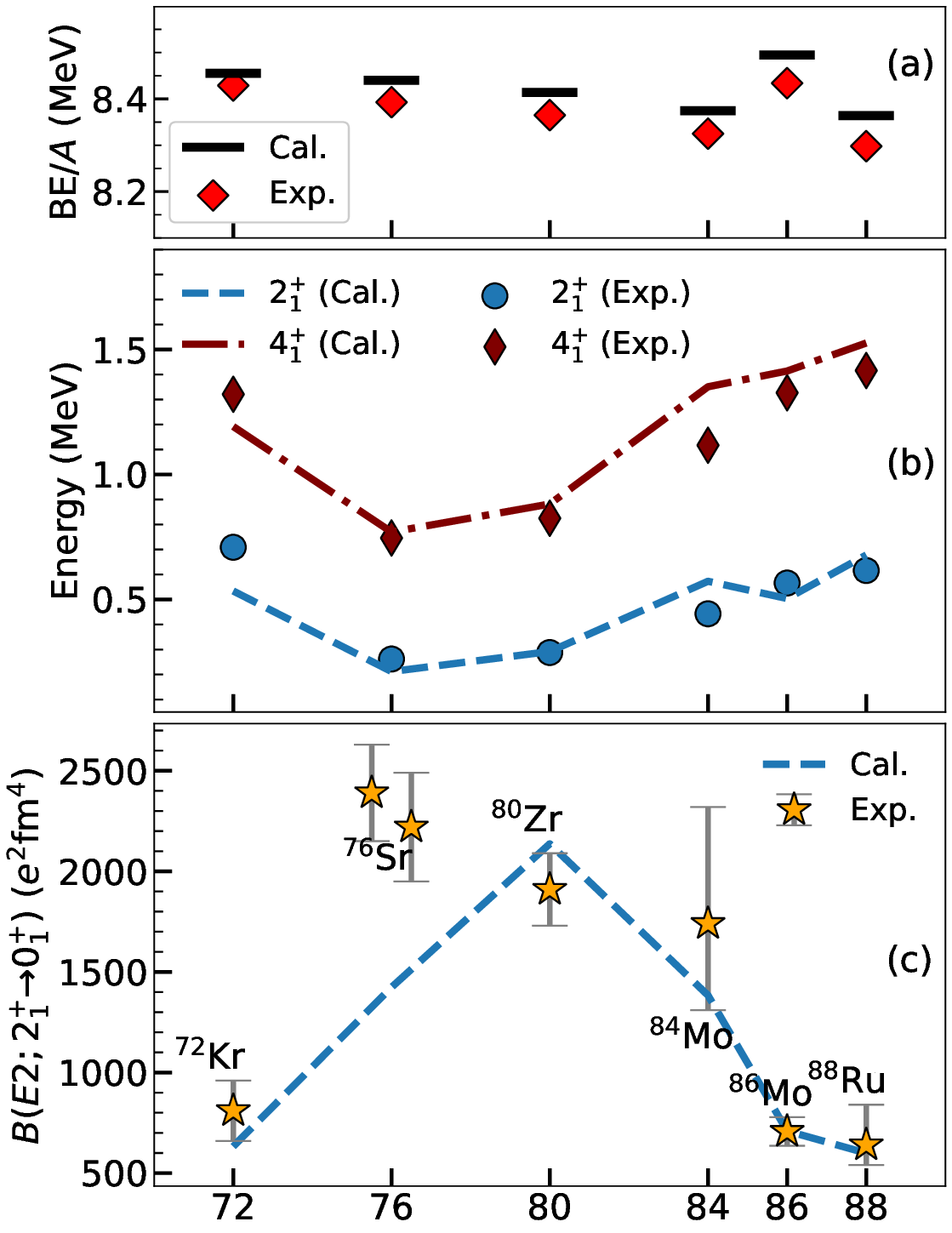}
  \caption{Calculated (a) binding energies per nucleon, (b) excitation energies of  $2_1^+$ and $4_1^+$ states, and (c) $B(E2;2_1^+\rightarrow0_1^+)$ values of the $N=Z$ nuclei from $^{72}$Kr to $^{88}$Ru, compared with experimental data. The values for $^{86}$Mo are also illustrated. Experimental data are taken from Refs.~\cite{Wang_2021,HamakerNP2021,PhysRevLett.124.152501,PhysRevLett.49.308,PhysRevC.85.041303,PhysRevLett.112.142502,HaNC2025,r6ns-ypw8,2dyn-q7wp,ENSDF}.\label{Fig2}}
\end{figure}

After configuration mixing, we examine the collective wave functions. The collective amplitudes of wave functions are constructed from the Hill-Wheeler-Griffin solutions in the natural basis~\cite{Ring1980,PhysRevC.79.044312,PhysRevC.81.064323}. Their squared amplitudes, normalized to unity, give the probability distributions of the PGCM states over the quadrupole coordinates $q=(Q_{20},Q_{22})$ (see details in the Supplemental Material and Refs.~\cite{Ring1980,PhysRevC.79.044312,PhysRevC.81.064323}). As shown in Figs.~\ref{Fig1}(c)-1(f), the $0^+_1$, $2^+_1$, $4^+_1$, and $6^+_1$ states all have their dominant weight near $Q_{20}\approx100$ $\mathrm{fm}^2$ and small $Q_{22}$. Little weight appears in the spherical or triaxial regions. The yrast band is therefore built on the same prolate intrinsic structure, rather than on a broad mixture of coexisting shapes.

The spectroscopic observables in Figs.~\ref{Fig1}(g) and~\ref{Fig1}(h) are consistent with this strongly prolate picture. The VS-IM-GCM calculation reproduces the observed low-lying rotational band and yields a ground-state energy that differs from the measured value by less than $0.6\%$. The calculated $B(E2;2^+_1\rightarrow0^+_1)=2137$ $e^2\mathrm{fm}^4$ is in good agreement with the measured $1910(180)$ $e^2\mathrm{fm}^4$. Omitting the induced two-body part of the evolved $E2$ operator
changes the calculated transition strength by
$141~e^2{\rm fm}^4$, about $6.6\%$ of the full result. This minor effect from the induced two-body $E2$ component is in accord with the in-medium generator-coordinate method (IM-GCM) result obtained with operators evolved in the multi-reference IMSRG~\cite{dfxs-41y3}.

Tetrahedral deformation has been proposed to explain the $\delta V_{pn}^{(3)}$ peak in $^{80}$Zr~\cite{89v1-lrb2}.
To describe its positive-parity yrast band and large $E2$ strength, we restrict the generator coordinates to $(Q_{20},Q_{22})$. Possible $Q_{32}$ correlations and their effects on local mass differences are not assessed here.


We now extend the same calculation from $^{80}$Zr to the even-even $N=Z$ sequence from $^{72}$Kr to $^{88}$Ru. Figure~\ref{Fig2} compares the calculated binding energies per nucleon, $2^+_1$, $4^+_1$ energies, and $B(E2;2^+_1\rightarrow0^+_1)$ values from $^{72}\mathrm{Kr}$ to $^{88}\mathrm{Ru}$, with $^{86}\mathrm{Mo}$ included as the $N=Z+2$ benchmark. Our calculation reproduces the binding energies within $0.8\%$. The calculation yields low excitation energies and large $E2$ strengths in $^{76}\mathrm{Sr}$, $^{80}\mathrm{Zr}$, and $^{84}\mathrm{Mo}$, followed by a rapid decline in collectivity toward $^{86}\mathrm{Mo}$ and $^{88}\mathrm{Ru}$. The overall trend is well reproduced, although the calculation underestimates the measured $B(E2;2_1^+\rightarrow0_1^+)$ in $^{76}$Sr. Uncertainties in the interaction~\cite{annurev:/content/journals/10.1146/annurev-nucl-101917-021120} and correlations omitted by the IMSRG(2)~\cite{PhysRevC.96.034324,HENDERSON2018468,PhysRevC.102.034320,PhysRevC.105.034333} and PGCM approximations~\cite{PhysRevC.96.054310,PhysRevC.98.064324,PhysRevC.110.054326,CAO2025140034} may contribute to this difference. Their effects are not separated here.

\begin{figure}[t]
  \includegraphics[width=\columnwidth]{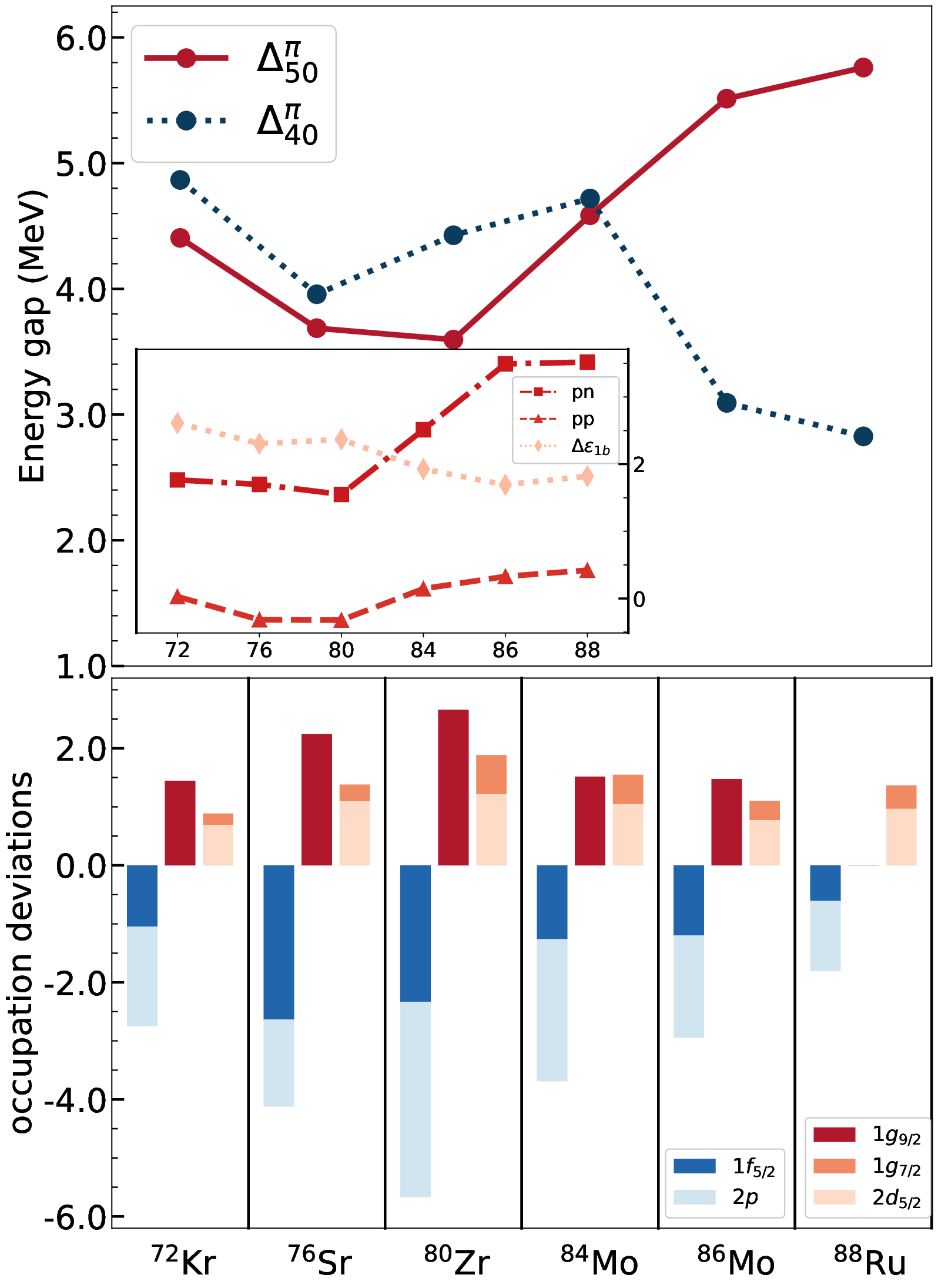}
  \caption{Proton shell evolution and cross-shell occupation deviations for $2_1^+$ states along the $N=Z$ line, with $^{86}$Mo included as an $N=Z+2$ benchmark. The upper panel shows the proton ESPE gaps across the $Z=40$ and $Z=50$ closures for $2_1^+$ states, defined as $\Delta^\pi_{40}=\epsilon^\pi_{1g_{9/2}}-\epsilon^\pi_{2p_{1/2}}$ and $\Delta^\pi_{50}=\epsilon^\pi_{2d_{5/2}}-\epsilon^\pi_{1g_{9/2}}$. The inset decomposes $\Delta^\pi_{50}$ into the one-body spacing $\Delta\varepsilon_{1b}$ between $2d_{5/2}$ and $1g_{9/2}$, the proton-proton monopole contribution, and the proton-neutron monopole contribution. The lower panel shows the stacked proton occupation deviations $\delta N^\pi_\mathbf{I}=\sum_{i\in \mathbf{I}}(\langle n^\pi_i\rangle-n^{\pi,0}_i)$ for $2_1^+$ states. Three columns denote the set $\mathbf{I}$ of orbits below the $Z=40$ gap, the $1g_{9/2}$ orbit between the $Z=40$ and $Z=50$ gaps, the $2d_{5/2}$ and the $1g_{7/2}$ orbits above the $Z=50$ gap, respectively. Each column presents the detailed occupation deviations for specific orbits. Note that ``$2p$'' denotes the total proton deviations of $2p_{1/2}$ and $2p_{3/2}$ orbits.}
  \label{Fig3}
\end{figure}

To unveil the connection between shell evolution and rapid changes in quadrupole collectivity, Fig.~\ref{Fig3} compares the proton ESPE gaps with the corresponding occupation rearrangements. We define the ESPE gap at $Z=40$ and $Z=50$ as $\Delta^\pi_{40}=\epsilon^\pi_{1g_{9/2}}-\epsilon^\pi_{2p_{1/2}}$ and $\Delta^\pi_{50}=\epsilon^\pi_{2d_{5/2}}-\epsilon^\pi_{1g_{9/2}}$, where $\epsilon^{\pi}_i$ denotes the ESPE of orbit $i$ given by 
\begin{equation}
\epsilon^{\pi}_{i}
=
\varepsilon^{\pi}_{1b,i}
+
\sum_j \langle n_j^\pi\rangle \, \bar V^{\pi\pi}_{ij}
+
\sum_j \langle n_j^\nu \rangle\, \bar V^{\pi\nu}_{ij}.
\end{equation}
$\varepsilon_{1b,i}$ denotes the one-body term of the orbit $i$ of the VS-IMSRG-evolved Hamiltonian, while $\bar V^{\tau\tau^{\prime}}_{ij}$ is the angular-momentum averaged two-body monopole matrix element defined in Refs.~\cite{PhysRevLett.87.082502,PhysRevLett.95.232502}, which is extracted from the same Hamiltonian. $\langle n_j^\tau\rangle$ is the occupation number of the orbit $j$ given by the full VS-IM-GCM calculations. The $Z=40$ ESPE gap remains open in $^{76}$Sr and $^{80}$Zr and does not follow the $E2$ systematics. By contrast, the $Z=50$ $1g_{9/2}$--$2d_{5/2}$ gap is reduced in these nuclei. Most of this variation comes from the proton-neutron monopole contribution (see inset of Fig.~\ref{Fig3}). The evolution of ESPEs characterizes the interplay of shell structure and quadrupole correlations. 

To separate cross-shell excitations from naive filling of single-particle levels in order of increasing energies, we define the proton occupation deviations $\delta N^\pi_\mathbf{I}=\sum_{i\in \mathbf{I}}
(\langle n^\pi_i\rangle-n^{\pi,0}
_i)$, where $n^{\pi,0}_i$ is obtained from filling the single-proton levels in order of increasing energies $\varepsilon^{\pi}_{1b}$, and $\mathbf{I}$ labels a group of orbits. Negative deviations indicate the $pf$ holes. Positive deviations indicate the protons promoted to the 1$g$ and 2$d_{5/2}$ orbits. Note that $1g_{9/2}$ and $2d_{5/2}$ orbits form a quasi-SU(3) pair with $\Delta j=\Delta\ell=2$, which favors strong quadrupole correlations~\cite{PhysRevC.52.R1741,PhysRevC.92.024320}. As shown in the lower panel of Fig.~\ref{Fig3}, the reduced $1g_{9/2}$--$2d_{5/2}$ spacing in $^{76}$Sr and $^{80}$Zr favors simultaneously enhanced occupation of the $1g_{9/2}$ and $2d_{5/2}$ orbits, while more holes develop in the $pf$ shell. These occupations are consistent with coupling the pseudo-SU(3) holes to the quasi-SU(3) $1g_{9/2}$--$2d_{5/2}$ particles. The $1g_{9/2}$--$2d_{5/2}$ spacing exceeds $5.5$ MeV toward $^{86}\mathrm{Mo}$ and $^{88}\mathrm{Ru}$, and the number of $pf$-shell holes decreases in $^{88}\mathrm{Ru}$, weakening this coherent quadrupole coupling. Note that the occupations of the other upper-shell orbits are also appreciable in $^{80}$Zr (see Supplemental Material), so this interpretation does not imply confinement to a pure pseudo-/quasi-SU(3) subspace. 

\begin{figure}[t]
  \includegraphics[width=\columnwidth]{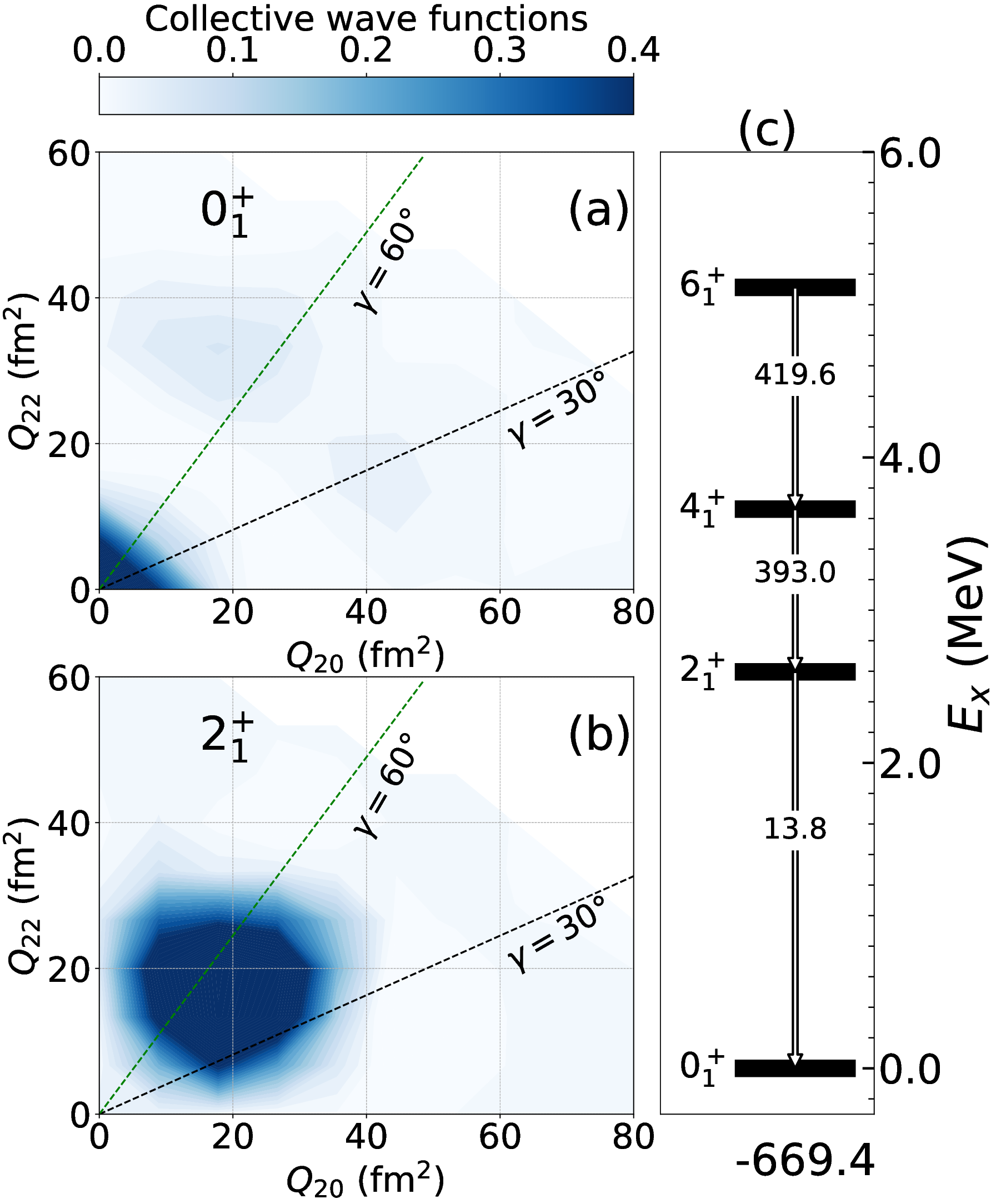}
    \caption{Distributions of collective wave functions of the $0_1^+$ and $2_1^+$ states, low-lying spectrum, and the $B(E2\downarrow)$ values (in $e^2\mathrm{fm}^4$) of $^{80}$Zr using the same effective Hamiltonian and the consistently evolved $E2$ operator as those used in Fig.~\ref{Fig1}, but obtained with the restricted VAPNP basis in which occupations of both proton and neutron $2d_{5/2}$ orbits are constrained to zero.}
  \label{Fig4}
\end{figure}

Similar neutron shell evolution and occupation deviations are presented in the Supplemental Material, indicating that this mechanism is largely isospin symmetric. The resulting pseudo-SU(3) hole and quasi-SU(3) particle picture is consistent with the earlier LSSM~\cite{KANEKO2021136286,HaNC2025} and DNO-SM studies~\cite{HaNC2025,r6ns-ypw8} of this region, but here it emerges from the chiral NN+3N force that governs the shell structure evolution, orbital occupations, and quadrupole correlations, without imposing any SU(3)-based truncation.

To test the role of $2d_{5/2}$, we repeat the PGCM calculation
for $^{80}$Zr with intrinsic states constrained to zero proton
and neutron occupation of this orbit. The Hamiltonian and evolved
$E2$ operator remain unchanged. As shown in Fig.~\ref{Fig4}, the collective distribution of the $0_1^+$ state becomes nearly spherical, and the $2_1^+$ state becomes weakly deformed. The yrast spectrum is stretched, and the $E2$ strength is significantly suppressed from $2137$ to $13.8~e^2\,\mathrm{fm}^4$. This dramatic change supports the importance of $2d_{5/2}$ for describing the enhancement of quadrupole collectivity. The restriction removes all correlations involving this orbit and does not isolate a particular monopole term or a unique SU(3) component.

{\it Summary.}--
Starting from the same chiral NN+3N interaction, the VS-IM-GCM framework reproduces both the emergence of strong quadrupole collectivity near $^{80}\mathrm{Zr}$ and its rapid reduction toward $^{86}\mathrm{Mo}$ and $^{88}\mathrm{Ru}$, without phenomenological adjustments in the $A\approx80$ region or empirical effective charges. The $N,Z=40$ ESPE gap does not collapse and does not track the measured $E2$ strength. Instead, the proton-neutron monopole interaction contributes less to the $1g_{9/2}$--$2d_{5/2}$ ESPE gap in $^{76}$Sr and $^{80}$Zr. It results in a concurrent increase in the occupation of the $1g_{9/2}$ and $2d_{5/2}$ orbits, allowing pseudo-SU(3) $pf$ holes and quasi-SU(3) particles to be strongly coupled in low-lying states. The widening of this spacing in $^{86}$Mo and the reduction of $pf$ holes in $^{88}$Ru remove this favorable combination and mark the loss of collectivity. The calculation thus connects the abrupt onset and weakening of quadrupole collectivity directly to shell evolution rooted in microscopic nuclear forces derived from the chiral effective field theory. It marks a significant step toward the accurate \textit{ab initio} description of deformed heavy nuclei characterized by  strong collectivity.

{\it Acknowledgments.}--
The authors would like to thank J. M. Yao and J. G. Li for fruitful discussions and useful comments. We thank T. Miyagi for the \texttt{NuHamil} code~\cite{Miyagi2023} which was used to generate chiral EFT matrix elements, and R. Stroberg for \texttt{IMSRG++}~\cite{Stroberg_IMSRG} which was used to perform the VS-IMSRG decoupling. The authors also thank B. Bally and T. R. Rodr\'{i}guez for making the \texttt{TAURUS} codes available~\cite{Bally2021,Bally2024}, on which our PGCM implementation is based. This material is based on the work supported by the National Natural Science Foundation of China under Grant No. 12275369.  

\bibliography{ISIoI}

\section{Supplemental material for ``\textit{Ab Initio} Emergence and Collapse of Nuclear Collectivity near \textit{N}=\textit{Z}=40''}







This Supplemental Material presents numerical checks and additional
structural results for the even-even $N=Z$ nuclei near $A=80$
and the neighboring $N=Z+2$ nucleus $^{86}$Mo.

\subsection{The convergence check of the VS-IM-GCM calculation}

\begin{figure}[t]  \includegraphics[width=\columnwidth]{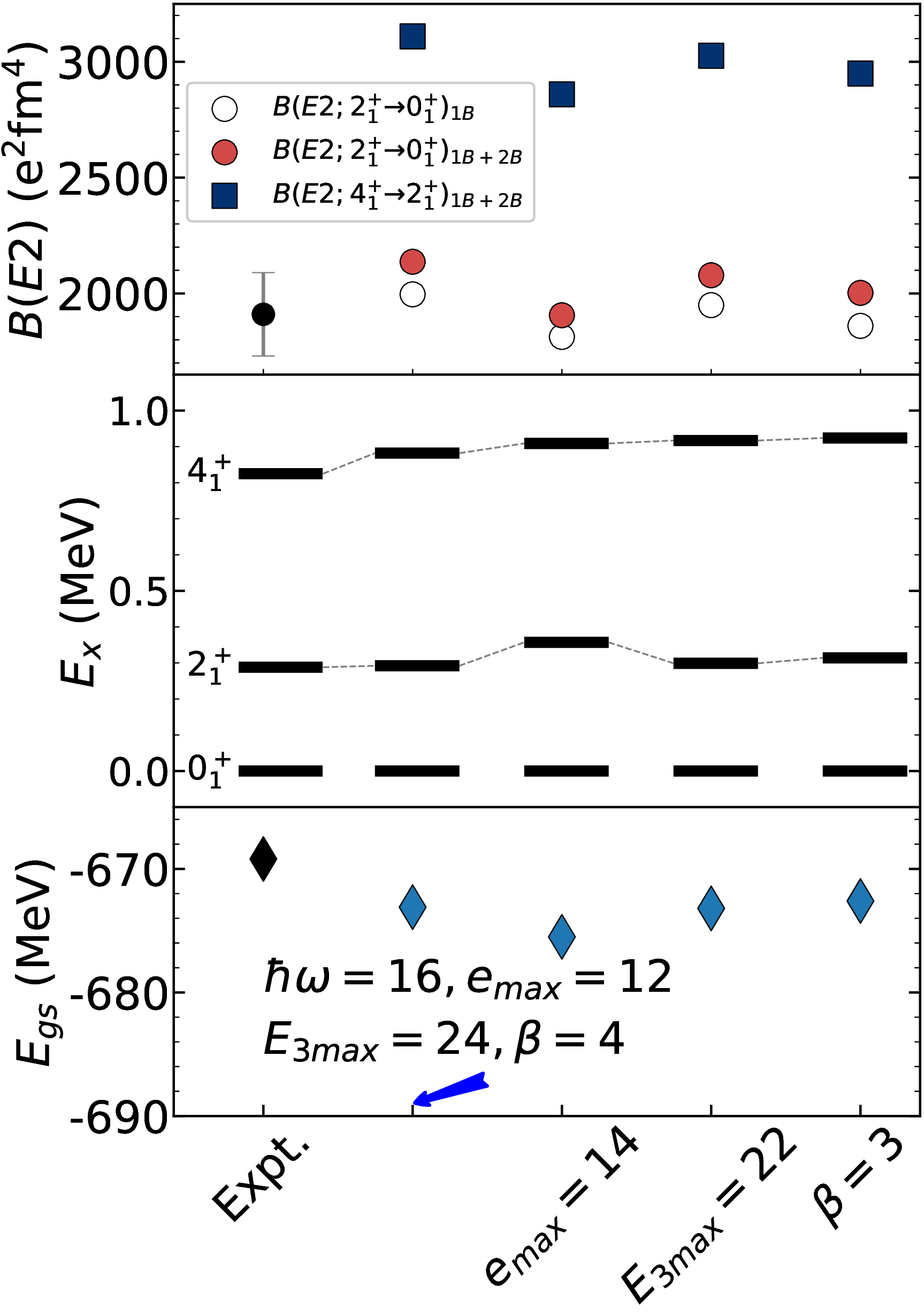}
  \caption{$B(E2;2_{1}^{+}\to0_{1}^{+})$ and $B(E2;4_{1}^{+}\to2_{1}^{+})$ values (top), low-lying spectra (middle) and the ground state energies (bottom panel) of $^{80}$Zr given by our VS-IM-GCM calculations with the 1.8/2.0(EM) NN+3N chiral interaction. The experimental ground state energy is given by the mass measurement in Ref.~\cite{HamakerNP2021}. The third to fifth columns show the weak dependence of our main results on the single-particle basis truncation $e_{\mathrm{max}}$, the $E_{\mathrm{3max}}$ restriction of 3N matrix elements, and the choice of $\beta$.}
  \label{convergence test}
\end{figure}

We first consider $^{80}$Zr to test the convergence behavior of our VS-IM-GCM calculation, focusing on the uncertainties arising from the truncation in the single-particle basis $e_{\mathrm{max}}$, the restriction of the three-body matrix elements $E_{3\mathrm{max}}$, and the choice of scaling factor $\beta$ for the Gl\"ockner-Lawson term~\cite{GLOECKNER1974313,PhysRevC.102.034320}. Figure~\ref{convergence test} shows our VS-IM-GCM results for the $0_1^+$ ground-state energies, the low-lying spectra, as well as the $B(E2;2_1^+\rightarrow0_1^+)$ and $B(E2;4_1^+\rightarrow2_1^+)$ values based on the 1.8/2.0(EM) chiral interaction, with different theoretical choices. Our main results for all the investigated $A\approx80$, $N\approx Z$ nuclei use $e_{\mathrm{max}}=12$, $E_{3\mathrm{max}} = 24$, $\hbar\omega = 16$ MeV. For $^{80}$Zr, the main result is obtained with $\beta = 4$. In Fig.~\ref{convergence test}, we show in detail that the dependence on these theoretical choices is weak. We use this variation to demonstrate the robustness of our VS-IM-GCM results within the large $N(Z) = 28$ to 82 valence space. These variations do not constitute a full uncertainty estimate, as the full contributions from Hamiltonian uncertainty and many-body uncertainty are not taken into account. For the $B(E2;2_1^+\rightarrow0_1^+)$ values, we show the small contributions of the two-body part of the $E2$ operator induced by the VS-IMSRG evolution.

\subsection{The single-particle energies calculated with the VS-IMSRG-evolved Hamiltonian}

\begin{figure*}[ht!]
    \centering
    \includegraphics[width=\textwidth]{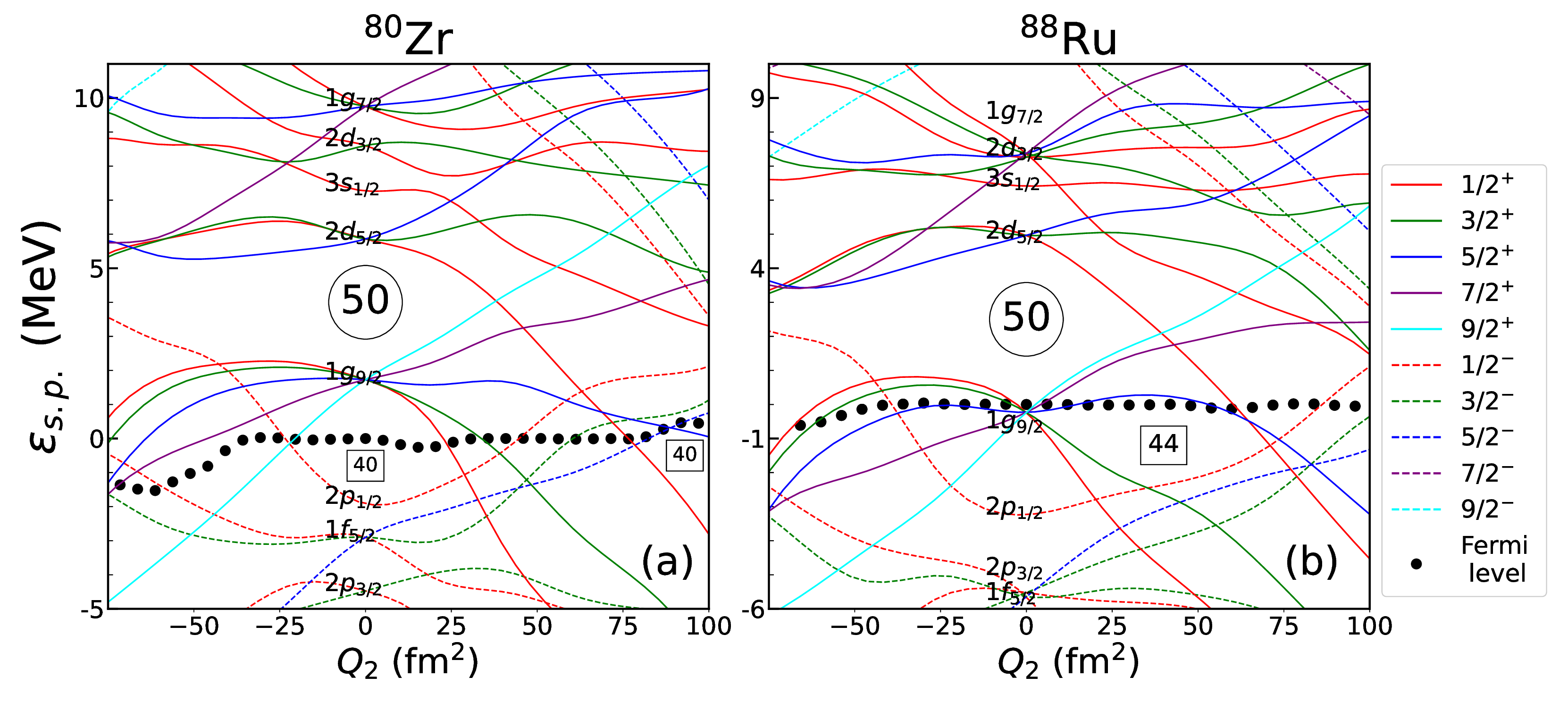}
    \caption{Single-particle energies for protons of $^{80}$Zr (a) and $^{88}$Ru (b) as a function of the quadrupole moment $Q_{2}$, calculated with the effective VS-IMSRG Hamiltonian based on the 1.8/2.0(EM) chiral interaction. The Fermi level is denoted by black dots.\label{Nilsson}}
\end{figure*}

In Fig.~\ref{Nilsson}, we display the single-proton energies for $^{80}$Zr and $^{88}$Ru calculated with the effective VS-IMSRG Hamiltonian based on the 1.8/2.0(EM) chiral interaction, as a function of the quadrupole moment $Q_{2}$. The single-particle energies are obtained by diagonalizing the Hartree--Fock single-particle Hamiltonian $h$, constructed from the intrinsic density matrix of the converged variation-after-particle-number-projection (VAPNP) solution. The single-neutron energies follow a similar pattern with energies about 11 MeV lower. For $^{80}$Zr, with increasing quadrupole moment, the $2p_{1/2}$ and a part of $1f_{5/2}$ levels are depopulated while a part of $1g_{9/2}$ and $2d_{5/2}$ orbits drop below the Fermi level. This structure favors the formation of both strongly deformed $Z=40$ and $N=40$ shell gaps at $Q_{2}\approx100\text{ fm}^2$. For $^{88}$Ru, a weakly deformed $Z(N)=44$ gap appears at $Q_{2}\approx 45\text{ fm}^2$, because the lower part of $1g_{9/2}$ orbits enter below the Fermi energy of $Z(N)=44$.

\subsection{The distribution of collective wave functions\label{collective wave functions}}

\begin{figure*} [t]
    \centering
    \includegraphics[width=\textwidth]{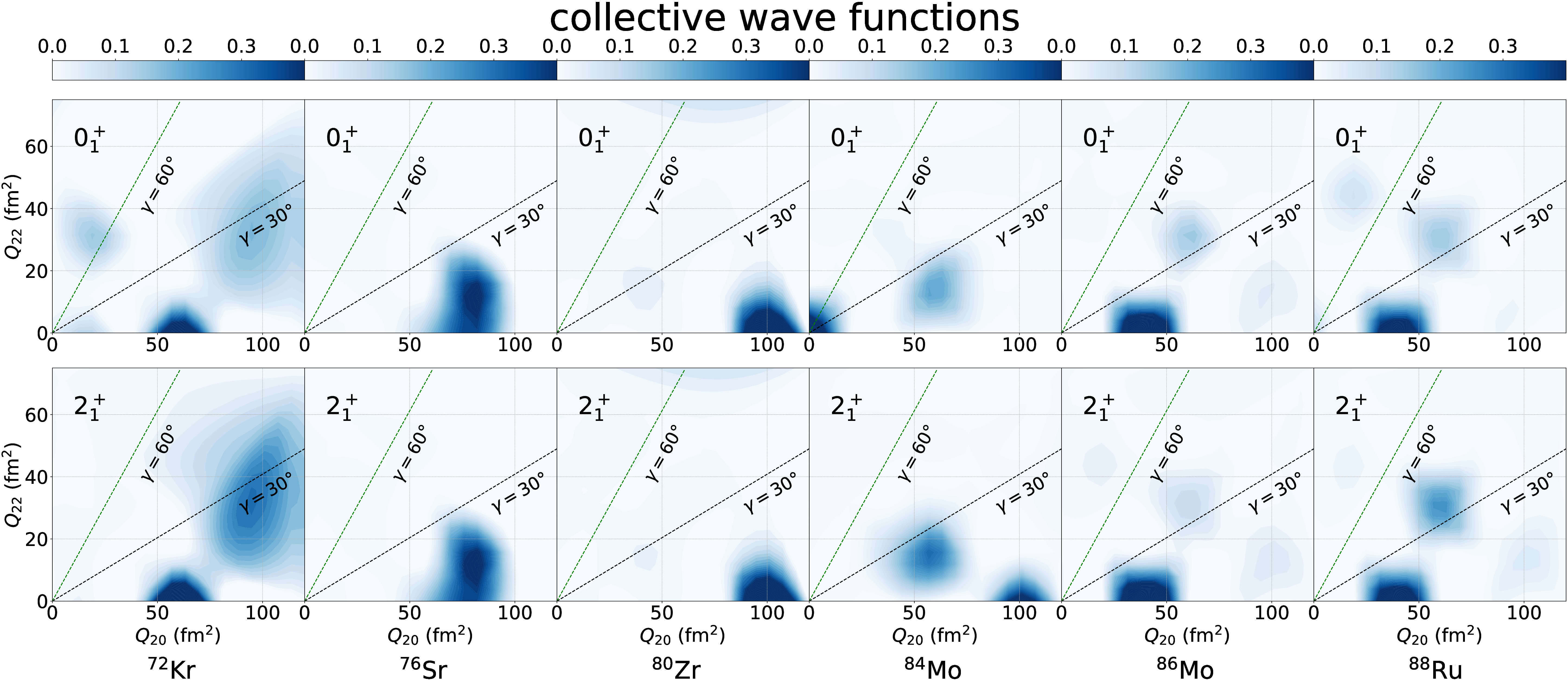}
    \caption{The distributions of collective wave functions corresponding to $0_1^+$ (upper) and $2_1^+$ (lower) states in $^{72}$Kr, $^{76}$Sr, $^{80}$Zr, $^{84}$Mo, $^{86}$Mo, and $^{88}$Ru, respectively.}
    \label{collective wave functions}
\end{figure*}

The PGCM many-body wave functions are written as
\begin{equation}\label{PGCMwf}
|\Psi_{\sigma}^{JNZ}\rangle
=\sum_{q,K} f_{q,\sigma}^{JK}|JMK;NZ;q\rangle,
\end{equation}
where $|JMK;NZ;q\rangle
\equiv \hat {\mathcal{P}}_{MK}^{J}\hat {\mathcal{P}}^{N}\hat {\mathcal{P}}^{Z}|\Phi(q)\rangle$, and the collective coordinate $q=(Q_{20},Q_{22})$. The intrinsic states $|\Phi(q)\rangle$ constrained to different $q$ constitute a nonorthogonal intrinsic basis set $\it\Gamma$.
The weight functions $f_{q,\sigma}^{JK}$ can be determined by solving the Hill--Wheeler--Griffin (HWG) equation 
\begin{equation}
\sum_{q^{\prime}K^{\prime}}\left[
\mathcal H^{J}_{KK^{\prime}}(q,q^{\prime})
-E_{\sigma}^{JNZ}\mathcal N^{J}_{KK^{\prime}}(q,q^{\prime})
\right]f_{q^{\prime},\sigma}^{JK^{\prime}}=0,
\end{equation}
where norm and Hamiltonian kernels are
\begin{equation}
\begin{aligned}
\mathcal N^{J}_{KK^{\prime}}(q,q^{\prime})
&=\langle JMK;NZ;q|JMK^{\prime};NZ;q^{\prime}\rangle,\\
\mathcal H^{J}_{KK^{\prime}}(q,q^{\prime})
&=\langle JMK;NZ;q|\hat H|JMK^{\prime};NZ;q^{\prime}\rangle.
\end{aligned}
\end{equation}

Because $|JMK;NZ;q\rangle$ are nonorthogonal, the
coefficients $f_{q,\sigma}^{JK}$ are \textit{not} probability amplitudes. In the PGCM framework, the HWG equation is essentially a generalized eigenvalue problem, which should be transformed into a Schrödinger-like equation first. We therefore extract an orthonormal and linearly independent set of states, the so-called \textit{natural basis}, from the original set $\it\Gamma$ of nonorthogonal intrinsic states. Following the natural-basis constructions of Refs.~\cite{Ring1980,PhysRevC.79.044312,PhysRevC.81.064323}, we first solve the eigenvalue problem of the norm kernel,
\begin{equation}
\begin{aligned}
&\sum_{q^{\prime}K^{\prime}}\mathcal N_{K K^{\prime}}^{J}(q,q^{\prime})
u_{\kappa}^{J}(q^{\prime},K^{\prime})
=n_{\kappa}^{J}u_{\kappa}^{J}(q,K),\\
&\sum_{qK}u_{\kappa}^{J*}(q,K)u_{\kappa^{\prime}}^{J}(q,K)
=\delta_{\kappa\kappa^{\prime}}.
\end{aligned}
\end{equation}
Here $\kappa$ labels a retained eigenvector with $n_{\kappa}^{J}>0$.  With
$U_{qK,\kappa}^{J}\equiv u_{\kappa}^{J}(q,K)$ and
$n_{\kappa\kappa^{\prime}}^{J}\equiv
n_{\kappa}^{J}\delta_{\kappa\kappa^{\prime}}$, the
spectral representation of the norm kernel and its positive square root are
\begin{equation}
\begin{aligned}
&\mathcal N^{J}
=U^{J}n^{J}U^{J\dagger},\\
&\left(\mathcal N^{J}\right)^{1/2}
=U^{J}\left(n^{J}\right)^{1/2}U^{J\dagger},\\
&\left[\left(\mathcal N^{J}\right)^{1/2}\right]_{K K^{\prime}}
(q,q^{\prime})
=\sum_{\kappa}u_{\kappa}^{J}(q,K)\sqrt{n_{\kappa}^{J}}
u_{\kappa}^{J*}(q^{\prime},K^{\prime}).
\end{aligned}
\end{equation}
The associated natural states are constructed as
\begin{equation}
\begin{aligned}
&|\kappa;JMNZ\rangle
=\sum_{qK}\frac{u_{\kappa}^{J}(q,K)}{\sqrt{n_{\kappa}^{J}}}
|JMK;NZ;q\rangle,
\\
&\langle\kappa|\kappa^{\prime}\rangle=\frac{u_{\kappa}^{J\dagger}\mathcal{N}^{J}u_{\kappa^{\prime}}^{J}}
{\sqrt{n_{\kappa}^{J}n_{\kappa^{\prime}}^{J}}}
=\delta_{\kappa\kappa^{\prime}},
\end{aligned}
\end{equation}
which are orthonormal and define the ``collective'' subspace. Expanding the PGCM state in this orthonormal basis and comparing with
Eq.~(\ref{PGCMwf}) gives
\begin{equation}
\begin{aligned}
&|\Psi_{\sigma}^{JNZ}\rangle
=\sum_{\kappa}G_{\kappa,\sigma}^{J}|\kappa;JMNZ\rangle,\\
&f_{q,\sigma}^{JK}
=\sum_{\kappa}
\frac{u_{\kappa}^{J}(q,K)}{\sqrt{n_{\kappa}^{J}}}
G_{\kappa,\sigma}^{J},\\
&G_{\kappa,\sigma}^{J}
=\sqrt{n_{\kappa}^{J}}
\sum_{qK}u_{\kappa}^{J*}(q,K)f_{q,\sigma}^{JK},\\
&\sum_{\kappa}|G_{\kappa,\sigma}^{J}|^{2}=1.
\end{aligned}
\end{equation}
Mapping the normalized natural-basis amplitudes back to the
$(q,K)$ mesh then yields
\begin{equation}
\begin{aligned}
g_{\sigma}^{J}(q,K)
&\equiv\sum_{\kappa}u_{\kappa}^{J}(q,K)G_{\kappa,\sigma}^{J}\\
&=\sum_{q^{\prime}K^{\prime}}\sum_{\kappa}u_{\kappa}^{J}(q,K)
\sqrt{n_{\kappa}^{J}}\,u_{\kappa}^{J*}(q^{\prime},K^{\prime})
f_{q^{\prime},\sigma}^{J K^{\prime}}\\
&=\sum_{q^{\prime}K^{\prime}}
\left[\left(\mathcal N^{J}\right)^{1/2}\right]_{K K^{\prime}}
(q,q^{\prime})f_{q^{\prime},\sigma}^{J K^{\prime}} .
\end{aligned}
\end{equation}
Thus $g_{\sigma}^{J}(q,K)$, rather than $f_{q,\sigma}^{JK}$, is the
normalized collective amplitude on the generator-coordinate mesh. The distribution of collective
wave functions and its normalization are
\begin{equation}
\begin{aligned}
&P_{\sigma}^{J}(q)
=\sum_{K}|g_{\sigma}^{J}(q,K)|^{2},\\
&\sum_q P_{\sigma}^{J}(q)
=\sum_{qK}|g_{\sigma}^{J}(q,K)|^{2}=f_{\sigma}^{J\dagger}\mathcal N^{J}f_{\sigma}^{J}=1.
\end{aligned}
\end{equation}

Figure~\ref{collective wave functions} shows the distribution of collective wave functions $P_{\sigma}^{J}(q)$ for the $0_1^+$ and $2_1^+$ states of investigated $A\approx80$, $N\approx Z$ nuclei. We use the convention 
$\gamma=\arctan(\sqrt{2}Q_{22}/Q_{20})$ here. Thus, $\gamma\approx 0^{\circ}$ and $60^{\circ}$ correspond to the prolate and oblate shapes, respectively, whereas intermediate $\gamma$ values describe triaxially deformed configurations. In $^{72}$Kr, the weight is broadly distributed and shared by several sectors of the $(Q_{20},Q_{22})$ plane, consistent with a
transitional nuclear system with strong shape mixing. The distributions become sharply localized at large $Q_{20}\approx100$ $\mathrm{fm}^2$ and small $Q_{22}$ in $^{76}$Sr and
$^{80}$Zr, which is compatible with the strongly deformed $Z(N)=40$ shell gap shown in panel (a) of Fig.~\ref{Nilsson}. Their similar distributions of $0_1^+$ and $2_1^+$ states identify a stable
well-deformed prolate structure for the ground-state band. In $^{84}$Mo, deformed configurations remain prominent in the $2_1^+$ state. The  broader distribution of the $0_1^+$ collective wave function, combined with the appearance of the near-spherical component, indicates stronger mixing of weakly and moderately deformed configurations. Toward $^{86}$Mo and $^{88}$Ru, the
dominant weight moves to smaller $Q_{20}$ and spreads over finite $Q_{22}$, which is in accord with the $Z(N)=44$ shell gap displayed in panel (b) of Fig.~\ref{Nilsson}. It shows reduced quadrupole deformation and increased $\gamma$ softness.
This loss of localized strongly deformed weight can explain the reduction of the calculated $B(E2;2_1^+\!\rightarrow0_1^+)$ values discussed in the main text.

\begin{figure*}[ht!] \includegraphics[width=0.8\textwidth]{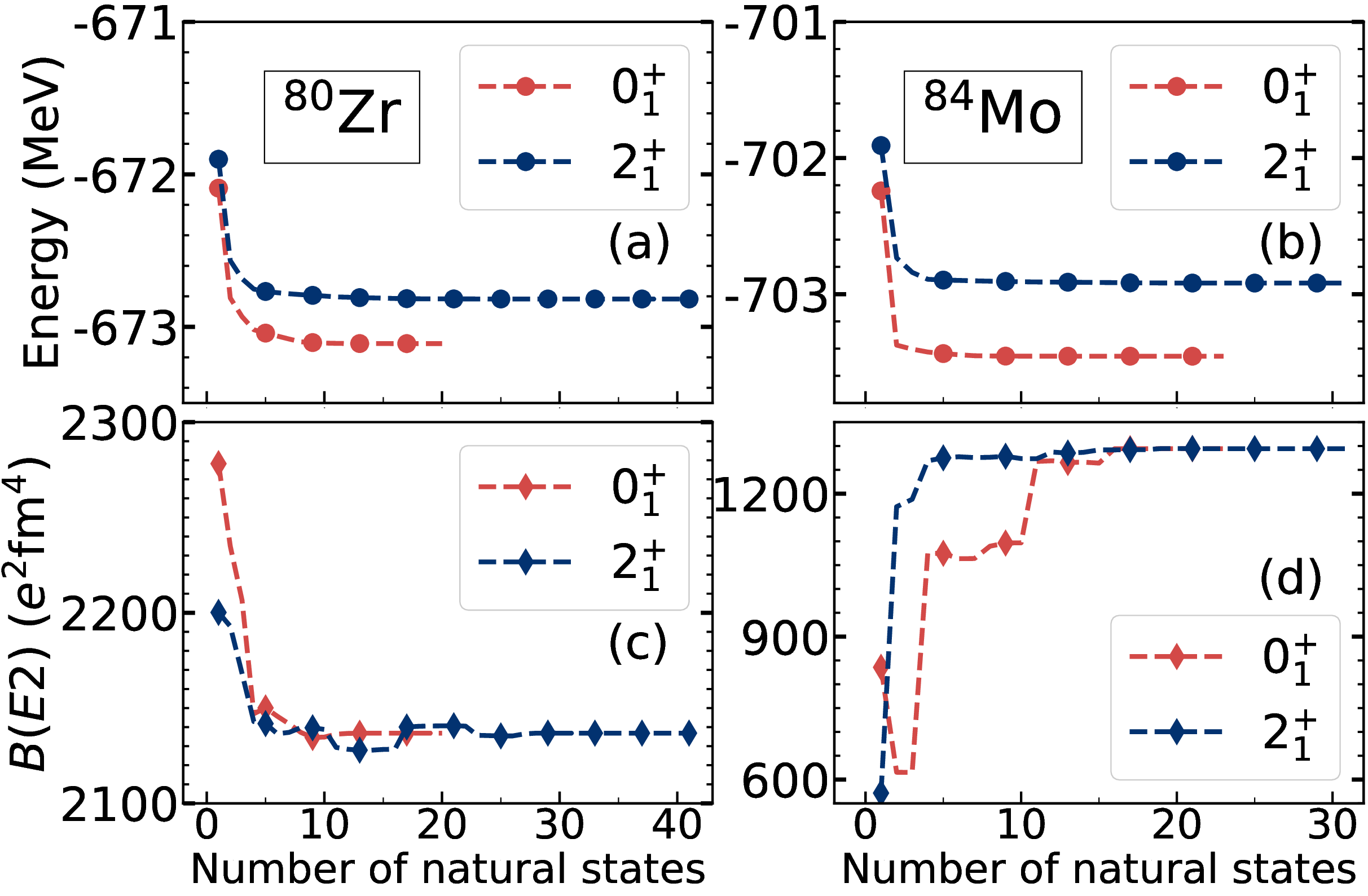}
  \caption{Calculated energies of $0_1^+$ and $2_1^+$ states [panels (a)--(b)] and the $B(E2;2_1^+\rightarrow0_1^+)$ [panels (c)--(d)] of $^{80}$Zr and $^{84}$Mo given by the VS-IM-GCM as a function of the dimension of the natural basis.}
  \label{convergence VAP states}
\end{figure*}

\subsection{Intrinsic-basis selection and PGCM convergence}

In practice, the variational solution of the PGCM may encounter numerical instabilities due to the overcompleteness and possible linear dependencies of the original intrinsic basis set $\it\Gamma$. To sufficiently capture the relevant collective correlations while removing the approximately linearly dependent configurations, we adopt an iterative prescription proposed in Ref.~\cite{Caurier1975} to truncate the intrinsic basis. The truncated basis set $\it\Gamma^{\prime}$ starts from the first intrinsic state $|\Phi(q_1)\rangle$ that minimizes the projected energy. We go over remaining states of the set $\it\Gamma$ and select the second intrinsic state $|\Phi(q_2)\rangle$ in such a way that the ground-state energy obtained from solving the HWG equation in the 2-dimensional space spanned by $|\Phi(q_1)\rangle$ and $|\Phi(q_2)\rangle$ is minimized. We move $|\Phi(q_2)\rangle$ from the original set $\it\Gamma$ to the truncated set $\it\Gamma^{\prime}$, and proceed in the same way in subsequent iterations. At the $k$th iteration, we select the $k$th state $|\Phi(q_k)\rangle$ which minimizes the ground-state energy given by solving the HWG equation in the $k$-dimensional space spanned by $\it\Gamma^{\prime}$ at the ($k-1$)th iteration and $|\Phi(q_k)\rangle$. Also, the intrinsic state is retained only when the eigenvalues of the corresponding norm kernel remain above a prescribed numerical threshold, ensuring linear independence. This procedure is continued until no further intrinsic state can be included in $\it\Gamma^{\prime}$ without violating the linear-independence criterion. For $J=0$, the final natural-basis dimension equals the number of selected intrinsic states. It provides an energetically truncated and numerically stable basis for the PGCM calculation.

Figs.~\ref{convergence VAP states} (a) and~\ref{convergence VAP states} (b) show the convergence of the PGCM $0_1^+$ and $2_1^+$ energies in $^{80}$Zr and $^{84}$Mo as the selected intrinsic basis is enlarged. The horizontal axis gives the number of retained natural states for each angular momentum. Both energies stabilize rapidly.
Energy convergence alone does not ensure convergence of transition matrix elements. We therefore examine $B(E2;2_1^+\rightarrow0_1^+)$ separately in Figs.~\ref{convergence VAP states} (c) and~\ref{convergence VAP states} (d). For each curve, the natural basis for the indicated state is enlarged while the other wave function is held fixed at the solution obtained in its largest retained basis. In $^{84}$Mo, the transition strength remains sensitive to the $0_1^+$ basis size after the energy has nearly stabilized. At larger basis dimensions, both sequences exhibit stable plateaus in each nucleus.

\begin{figure}[ht!]
  \includegraphics[width=\columnwidth]{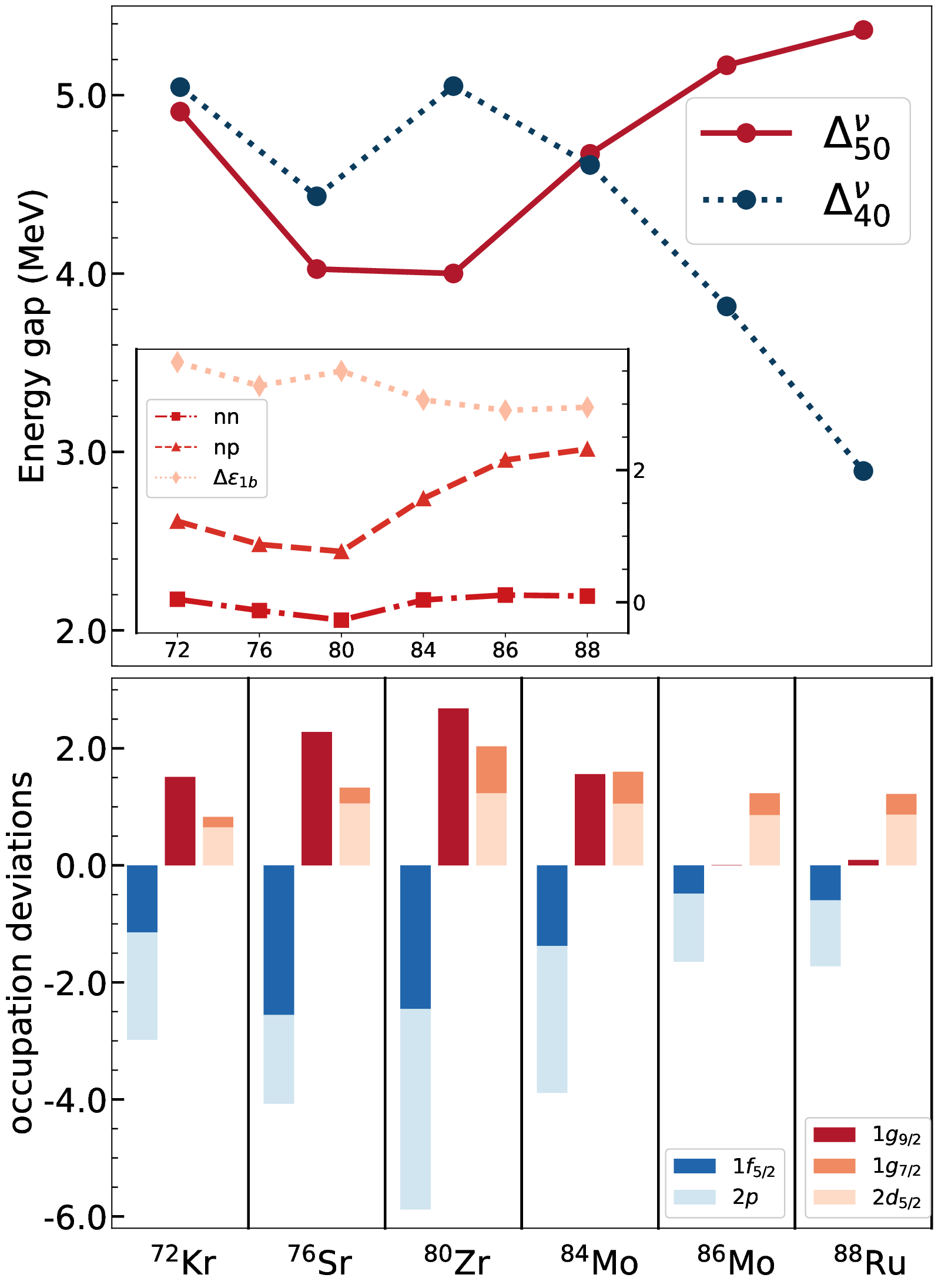}
  \caption{Neutron shell evolution and cross-shell occupation deviations of nuclei near $N=Z=40$. The upper panel shows the neutron ESPE gaps across the $N=40$ and $N=50$ closures for $2_1^+$ states, defined as $\Delta^\nu_{40}=\epsilon^\nu_{1g_{9/2}}-\epsilon^\nu_{2p_{1/2}}$ and $\Delta^\nu_{50}=\epsilon^\nu_{2d_{5/2}}-\epsilon^\nu_{1g_{9/2}}$. The inset decomposes $\Delta^\nu_{50}$ into the one-body spacing $\Delta\varepsilon_{1b}$ between $2d_{5/2}$ and $1g_{9/2}$, the neutron-neutron monopole contribution, and the neutron-proton monopole contribution. The lower panel shows the stacked neutron occupation deviations $\delta N^\nu_\mathbf{I}=\sum_{i\in \mathbf{I}}(\langle n^\nu_i\rangle-n^{\nu,0}_i)$ for $2_1^+$ states. $\mathbf{I}$ denotes a group of orbits below the $N=40$ gap, the $1g_{9/2}$ orbit between the $N=40$ and 50 gaps, the $2d_{5/2}$ and the $1g_{7/2}$ orbits above the $N=50$ gap, respectively. Each column presents the detailed occupation deviations for specific orbits. Note that ``$2p$'' denotes the total neutron deviations of $2p_{1/2}$ and $2p_{3/2}$ orbits.}
  \label{neutron ESPEs and p-h excitations}
\end{figure}

\begin{figure*}[ht!] 
\includegraphics[width=0.9\textwidth]{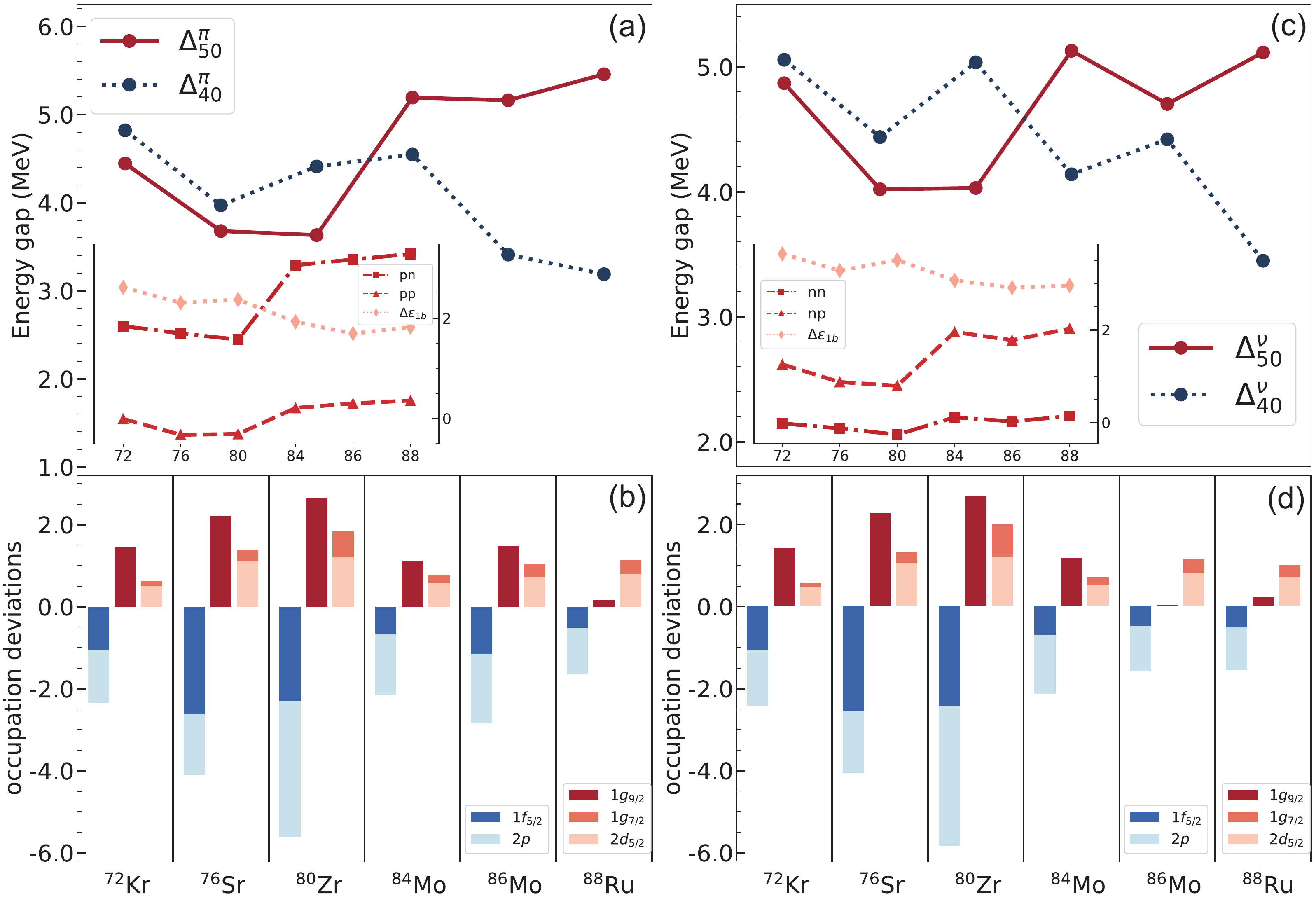}
  \caption{Similar to Fig.~\ref{neutron ESPEs and p-h excitations}, but evaluated using the $0_{1}^{+}$ states of nuclei near $N=Z=40$ for protons (a-b) and neutrons (c-d).}
  \label{ESPEs and p-h excitations 0+}
\end{figure*}

\subsection{ESPE gaps and occupation deviations\label{ESPE}}

Figure~\ref{neutron ESPEs and p-h excitations} is the neutron counterpart of Fig. 3 in the main text. The neutron results mirror the proton trends. $\Delta_{50}^{\nu}$ is smallest in
$^{76}$Sr and $^{80}$Zr, its variation is governed mainly by the neutron-proton monopole term, and the largest $pf$-shell depletion occurs in the same nuclei. Meanwhile, $\Delta_{40}^{\nu}$ decreases toward the upper end of the chain despite the loss of $E2$ strength. This is the neutron-side
confirmation of the isospin-symmetric shell evolution discussed in the main text. In $^{86}$Mo, the proton and neutron deviations use different independent-particle fillings because $Z=42$ and $N=44$; their absolute bar heights should not be compared directly.

Since the ESPEs [see Eq.~(1) in the main text] depend on the neutron and proton occupancies of the corresponding PGCM state, they are state-dependent. The ESPEs and occupation deviations presented in the main text are evaluated using the occupations of the $2_1^+$ state. Here we provide the neutron and proton ESPE gaps and occupation deviations evaluated using the $0_1^+$ states in Fig.~\ref{ESPEs and p-h excitations 0+}. We find that for $^{84}$Mo, the $\Delta_{50}$ gap obtained with $0_1^+$ occupation is considerably larger that that evaluated using $2_1^+$ state. This is consistent with the fact that the $0_1^+$ state of $^{84}$Mo strongly mixes with the nearly spherical configuration, while the $2_1^+$ state consists of well-deformed configurations. We therefore use the $2_1^+$ ESPEs and occupation deviations to characterize the cross-shell excitation associated with large deformation and strong quadrupole correlations. Note that this choice concerns only the structural analysis. The calculated $B(E2)$ involves both $0_1^+$ and $2_1^+$
PGCM states.

\begin{table}[t]
\caption{Occupations of valence-space orbits in VS-IM-GCM calculations for the $0^{+}_{1}$ and $2^{+}_{1}$ states. Proton and neutron occupancies are in the first and the second rows, respectively, for each state.\label{tab:occ}}
\renewcommand\arraystretch{0.8}
\begin{ruledtabular}
\begin{tabular}{lccccccccc}
 & $1f_{5/2}$ & $2p_{3/2}$ & $2p_{1/2}$ & $1g_{9/2}$ & $2d_{5/2}$ & $1g_{7/2}$ &$2d_{3/2}$ &$3s_{1/2}$&$1h_{11/2}$\\
\colrule
\multicolumn{10}{l}{$^{72}$Kr}\\
\multirow{2}{*}{$0^{+}_{1}$} & 2.94 & 2.11 & 0.61 & 1.44 & 0.50 & 0.13 &0.13 &0.09 & 0.05\\
             & 2.94 & 2.04 & 0.59 & 1.43 & 0.47 & 0.12 &0.11&0.08 & 0.22\\
\multirow{2}{*}{$2^{+}_{1}$} & 2.95 & 1.76 & 0.53 & 1.45 & 0.70 & 0.19 & 0.20 & 0.14 & 0.08 \\
                & 2.85 & 1.66 & 0.51 & 1.51 & 0.65 & 0.18 & 0.17& 0.12 &0.35\\
\colrule
\multicolumn{10}{l}{$^{76}$Sr}\\
\multirow{2}{*}{$0^{+}_{1}$} & 3.37 & 1.98 & 0.54 & 2.22 & 1.10 & 0.28 &0.28& 0.17& 0.05\\
            & 3.44 & 1.99 & 0.51 & 2.27 & 1.06 & 0.27 &0.25&0.16&0.06\\
\multirow{2}{*}{$2^{+}_{1}$} & 3.36 & 1.97 & 0.54 & 2.24 & 1.10 & 0.28 &0.28&0.17&0.06\\
            & 3.44 & 1.98 & 0.51 & 2.28 & 1.06 & 0.27 &0.25&0.16&0.06\\
\colrule
\multicolumn{10}{l}{$^{80}$Zr}\\
\multirow{2}{*}{$0^{+}_{1}$} & 3.69 & 2.03 & 0.65 & 2.66 & 1.20 & 0.65 & 0.72 & 0.31 & 0.07\\
          & 3.57 & 1.94 & 0.65 & 2.69 & 1.22 & 0.78 &0.79& 0.28&0.07\\
\multirow{2}{*}{$2^{+}_{1}$} & 3.67 & 2.01 & 0.65 & 2.66 & 1.22 & 0.67 &0.74&0.31&0.08 \\
          & 3.55 & 1.92 & 0.65 & 2.68 & 1.24 & 0.80 &0.81&0.28&0.08\\
\colrule
\multicolumn{10}{l}{$^{84}$Mo}\\
\multirow{2}{*}{$0^{+}_{1}$} & 5.34 & 3.40 & 1.12 & 3.10 & 0.58 & 0.20 &0.13&0.08&0.04 \\
          & 5.31 & 3.41 & 1.15 & 3.18 & 0.53 & 0.18 &0.11&0.07&0.04\\
\multirow{2}{*}{$2^{+}_{1}$} & 4.74 & 2.73 & 0.84 & 3.52 & 1.05 & 0.50 &0.41&0.16&0.05 \\
          & 4.62 & 2.66 & 0.83 & 3.56 & 1.06 & 0.54 &0.49&0.20&0.05\\
\colrule
\multicolumn{10}{l}{$^{86}$Mo}\\
\multirow{2}{*}{$0^{+}_{1}$} & 4.84 & 3.26 & 1.04 & 3.49 & 0.73 & 0.30 &0.16&0.12&0.05 \\
          & 5.53 & 3.52 & 1.37 & 4.04 & 0.82 & 0.34 &0.19&0.15&0.05\\
\multirow{2}{*}{$2^{+}_{1}$} & 4.80 & 3.22 & 1.03 & 3.48 & 0.78 & 0.32 &0.18&0.12&0.06 \\
          & 5.52 & 3.49 & 1.35 & 4.00 & 0.86 & 0.37 &0.21&0.15&0.06\\
\colrule
\multicolumn{10}{l}{$^{88}$Ru}\\
\multirow{2}{*}{$0^{+}_{1}$} & 5.48 & 3.54 & 1.34 & 4.17 & 0.80 & 0.33 &0.16&0.12&0.06\\
          & 5.49 & 3.58 & 1.38 & 4.24 & 0.72 & 0.29 &0.14&0.11&0.06\\
\multirow{2}{*}{$2^{+}_{1}$} & 5.39 & 3.50 & 1.30 & 4.00 & 0.97 & 0.40 &0.22&0.16&0.06 \\
          & 5.40 & 3.54 & 1.33 & 4.09 & 0.87 & 0.35 &0.20&0.15&0.06\\
\end{tabular}
\end{ruledtabular}
\end{table}

\subsection{Proton and neutron occupation numbers\label{occupancy}}

Table~\ref{tab:occ} lists the proton and neutron occupation numbers for each valence-space orbit. These values underlie the grouped occupation deviations shown in Fig.~3 of the main text and Figs.~\ref{neutron ESPEs and p-h excitations} and~\ref{ESPEs and p-h excitations 0+}. In $^{76}$Sr and $^{80}$Zr, the $0_1^+$ and $2_1^+$ values differ by no more than 0.02 for any valence-space orbit, consistent with a robust intrinsic structure for the ground-state band. In $^{80}$Zr, roughly 2.7 nucleons of each species occupy $1g_{9/2}$, and about three nucleons of each species occupy orbits above $N,Z=50$, indicating substantial multi-particle-multi-hole (\textit{mp--mh}) excitations across the $N,Z=40$ and 50 gaps. The strongest state dependence occurs in $^{84}$Mo. From the $0_1^+$ to the $2_1^+$ state, the total $pf$-shell occupation decreases by 1.55 for protons and 1.76 for neutrons, with corresponding increases in the $1g_{9/2}$ and higher-lying orbits. The similar proton and neutron occupancies are compatible with the isospin-symmetric cross-shell correlations discussed above.

\end{document}